\documentclass{article}

\usepackage{arxiv}

\usepackage[utf8]{inputenc} 
\usepackage[T1]{fontenc}    
\usepackage{lineno,hyperref}       
\usepackage{url}            
\usepackage{booktabs}       
\usepackage{amsfonts}       
\usepackage{nicefrac}       
\usepackage{microtype}      
\usepackage{graphicx}
\usepackage{natbib}
\usepackage{doi}
\usepackage{xcolor}
\usepackage{soul}
\usepackage{float}
\usepackage{amsmath,amsfonts}
\usepackage{algorithmic}
\usepackage{algorithm}
\usepackage{array}
\usepackage[caption=false,font=normalsize,labelfont=sf,textfont=sf]{subfig}
\usepackage{textcomp}
\usepackage{stfloats}
\usepackage{makecell}

\title{Fuel Consumption in Platoons: A Literature Review%
\thanks{This work has been submitted to the IEEE for possible publication. Copyright may be transferred without notice, after which this version may no longer be accessible.}}

\date{} 					

\author{ \href{}{Oumaima Barhoumi}\\
	Department of Electrical \\and Computer Engineering\\
	Concordia University\\
	Montral, Quebec, Canada \\
	\texttt{o\_barh@ece.concordia.ca} \\
	\And
\href{}{Ghazal Farhani} \\
	National Research Council Canada\\
    Automotive and Surface Transportation \\Research Centre\\
	London, Ontario, Canada\\
	\texttt{ghazal.farhani@nrc-cnrc.gc.ca} \\
    \And
\href{}{Taufiq Rahman} \\
	National Research Council Canada\\
    Automotive and Surface Transportation \\Research Centre\\
	London, Ontario, Canada \\
	\texttt{taufiq.rahman@nrc-cnrc.gc.ca} \\
    \And
\href{}{Mohamed H. Zaki} \\
	Department of Civil \\and Environmental Engineering\\
	Western University\\
	London, Ontario, Canada \\
	\texttt{mzaki9@uwo.ca} \\
    \And
\href{}{Sofiène Tahar} \\
	Department of Electrical \\and Computer Engineering\\
	Concordia University\\
	Montral, Quebec, Canada \\
	\texttt{tahar@ece.concordia.ca} \\
    \And
\href{}{Fadi Araji} \\
	Environment and Climate Change Canada\\
	Ottawa, Ontario, Canada \\
	\texttt{Fadi.Araji@ec.gc.ca}\\
    }

\renewcommand{\shorttitle}{Fuel Consumption in Platoons: A Literature Review}

\hypersetup{
pdftitle={Fuel Consumption in Platoons: A Literature Review},
pdfsubject={q-bio.NC, q-bio.QM},
pdfauthor={Oumaima Barhoumi, Ghazal Farhani, Taufiq Rahman, Mohamed H. Zaki, Sofiène Tahar and Fadi Araji},
pdfkeywords={Platooning, automated vehicles, fuel consumption, drag reduction, platoon instability, machine learning},
}

\begin{document}
\maketitle

\begin{abstract}
Platooning has emerged as a promising strategy for improving fuel efficiency in automated vehicle systems, with significant implications for reducing emissions and operational costs. While existing literature on vehicle platooning primarily focuses on individual aspects such as aerodynamic drag reduction or specific control strategies, this work takes a more comprehensive approach by bringing together a wide range of factors and components that contribute to fuel savings in platoons. In this literature review, we examine the impact of platooning on fuel consumption, highlighting the key components of platoon systems, the factors and actors influencing fuel savings, methods for estimating fuel use, and the effect of platoon instability on efficiency. Furthermore, we study the role of reduced aerodynamic drag, vehicle coordination, and the challenges posed by instability in real-world conditions. By compiling insights from recent studies, this work provides a comprehensive overview of the latest advancements in platooning technologies and highlights both the challenges and opportunities for future research to maximize fuel savings in real-world scenarios. 
\end{abstract}


\section{Introduction}
\label{intro}
In 2020, transportation emissions in Canada dropped sharply as the COVID-19 pandemic disrupted mobility nationwide. Despite this temporary decline, emissions have increased over the past few decades, rising by more than 30\% since 1990~\cite{statista_2024}. For instance, in 2021, transportation was responsible for 187.7 megatonnes (28.0\%) of overall greenhouse gas (GHG) emissions~\cite{statcan_commutes_2023}, where the largest proportion was from road transportation, including all types of vehicles and fuels. By 2023, transportation emissions reached 166 million metric tons of CO$_2$ (MtCO$_2$), marking a 1.3\% increase from 2022 
levels~\cite{statista_2024}. The recent rise in emissions highlights the ongoing challenge of curbing transportation-related GHG emissions, calling into question the effectiveness of existing mitigation policies. In this context, platooning~\cite{swaroop1994string} has been introduced by researchers and policymakers as a promising strategy to enhance road transportation efficiency and reduce greenhouse gas emissions. By enabling vehicles to travel in tightly coordinated convoys using connected and automated technologies, platooning minimizes aerodynamic drag, improves fuel economy, and supports broader climate goals in the transportation sector. 

In the 1990s, the concept of truck platooning was pioneered by the California PATH (Partners for Advanced Transit and Highways) program~\cite{rajamani2000demonstration}, a research initiative led by the University of California, Berkeley. The program sought to explore how automated vehicle coordination could enhance transportation efficiency. A major milestone came in 1997 \cite{rajamani2001experimental}, when PATH successfully demonstrated an 8-truck platoon on a California highway, utilizing early forms of automated control and vehicle-to-vehicle (V2V) communication to maintain tight, synchronized spacing. The core idea behind platooning was—and remains—that by driving in a closely coordinated line, trucks can significantly reduce aerodynamic drag, leading to improved fuel efficiency and better overall traffic flow. Platoons consist of strings of following-preceding automated vehicles traveling at the leader's speed and keeping a preset distance between one another. Platoons have been shown to significantly reduce fuel consumption for automated vehicles, primarily by lowering aerodynamic drag—especially at highway speeds. However, the magnitude of these savings depends on multiple interrelated factors, such as inter-vehicle distance, platoon size, vehicle type and load, and the consistency of driving behavior among vehicles. A critical additional factor is the information flow topology (IFT), which defines how communication is structured within the platoon. Effective IFTs enable quicker and smoother responses to changes in speed or acceleration, minimizing unnecessary throttle and brake inputs. External conditions such as traffic environment, and the emergence of platoon instabilities—like shockwaves from sudden speed changes—can further impact overall fuel efficiency.

There exist several surveys in the literature that examine different dimensions of vehicle platooning, particularly in relation to fuel efficiency. Among them, only a few literature reviews have addressed platooning in relation with fuel efficiency. For instance, in~\cite{zhou2016review}, the authors conduct a systematic review of fuel consumption models, identifying six primary factors that influence fuel economy: travel-related, weather-related, vehicle-related, roadway-related, traffic-related, and driver-related. In a similar context, the study in~\cite{li2017platoon} examines the existing literature from a network control perspective, aiming to assess platoon stability and its potential benefits in reducing fuel consumption. From an energy-saving standpoint,\cite{pi2023automotive} presents a survey of vehicle platoon strategies, focusing on aerodynamic optimization and vehicle speed control. More recently,\cite{zhang2024car} investigates the influence of car-following behavior on inter-vehicular interactions. Their work highlights the distinctions, complementarities, and overlaps among microscopic traffic flow and control models, based on their underlying principles and design logic. While the reviewed studies offer valuable insights into the factors influencing fuel consumption and the potential of vehicle platooning, none provide a fully integrated analysis that combines the diverse range of factors, components, and methods influencing platoon fuel savings. The studies in~\cite{zhou2016review} and~\cite{zhang2024car} lack a specific focus on platooning. For example,~\cite{zhou2016review} emphasizing on general fuel consumption models and~\cite{zhang2024car} centering on car-following behavior, hence limiting their relevance to platoon dynamics. Meanwhile, the work in~\cite{li2017platoon} and~\cite{pi2023automotive} adopt more focused scopes, concentrating on stability or optimization strategies. Specifically,~\cite{li2017platoon} concentrates on theoretical network control models aimed at ensuring platoon stability, with a strong emphasis on mathematical frameworks for maintaining consistent inter-vehicle spacing and velocity. However, this narrow focus limits its consideration of broader factors that influence fuel savings. Similarly,~\cite{pi2023automotive} primarily addresses aerodynamic drag reduction and speed optimization to enhance energy efficiency, but offers limited insight into how these strategies can be adapted to diverse platoon configurations or remain robust under external disturbances that may compromise platoon stability. A common observation for all above studies is that they tend to under-explore the critical issues of platoon instability and real-world implementation challenges, which restricts their practical deployment.

In contrast, the review we propose in this paper takes a comprehensive approach, integrating mathematical, experimental, and data-driven perspectives to comprehensively analyze how key factors affect fuel consumption in vehicle platooning. From a mathematical perspective, we delve into the models and equations that describe the complex interactions between vehicles, such as aerodynamic drag and vehicle dynamics. Experimental studies are reviewed to highlight real-world tests and simulations that validate theoretical findings, emphasizing how various platoon configurations and control methods affect fuel efficiency. Finally, we examine data-driven approaches, which leverage machine learning techniques to optimize platooning strategies and predict fuel savings under varying conditions. By integrating these three viewpoints, we aim to provide a comprehensive understanding of the key factors that influence fuel consumption in platooning, highlight existing gaps in the literature, and offer insights into future research directions. 

To conduct this survey, we adopted a systematic and multi-step methodology aimed at identifying and analyzing the most influential and relevant works in the field. Initially, we performed a comprehensive keyword-based search using Google Scholar, yielding approximately 290 papers to capture a broad spectrum of literature related to our topic. From this preliminary collection, we focused on highly cited papers and those published in top-tier conferences and journals, as these typically reflect a high level of impact and quality. We then employed a snowballing technique, closely examining the references cited within these core papers to uncover additional relevant studies that may not have surfaced in the initial search. This recursive approach allowed us to map out the foundational and emerging contributions in the field more thoroughly. In the final phase, we refined our selection by shortlisting papers based on their direct relevance to the scope of our survey, their novelty and contribution to the domain, the rigor and clarity of their methodology, and their citation count. This structured approach ensured a well-grounded and comprehensive survey of the state of the art.

The rest of the paper is organized as follows: Section~\ref{sec1} establishes the foundational concepts influencing platoon behavior, including vehicle dynamics, communication architectures, and spacing policies. These elements are critical, as fuel efficiency within platoons is strongly linked to behavioral factors and safety-constrained spacing strategies. Section~\ref{sec2} examines key contributors to fuel minimization in platoons, such as inter-vehicle aerodynamics, positional effects within the platoon, and additional parameters impacting fuel savings. In Section~\ref{sec3}, we review various fuel consumption models—analytical, numerical, and experimental—developed specifically for platoon scenarios. Section~\ref{sec4} addresses the important issue of how instability within the platoon affects fuel consumption. Finally, Section~\ref{sec5} identifies gaps in the current literature and outlines potential directions for future research. We conclude this work in Section~\ref{sec6}.

\section{Platoon Components}
\label{sec1}
\begin{figure}[htb!]
        \centering
        \includegraphics[width=0.7\linewidth]{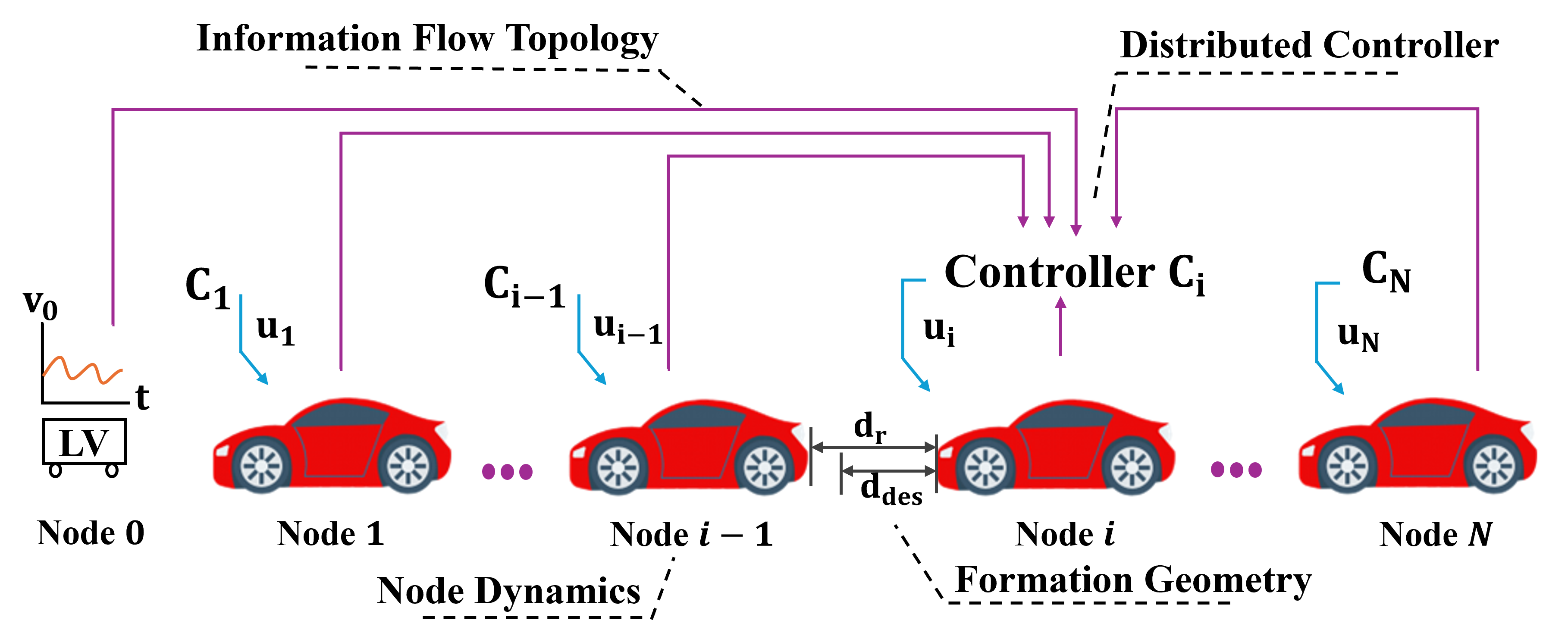} 
        \caption{Four major components of a platoon: 1) node dynamics, 2) information flow topology, 3) distributed controller, 4) geometry formation; where $d_{r}$ is the actual relative distance, $d_{des}$ is the desired distance, $u_{i}$ is the control signal for \textit{i}-th vehicle, and C denotes the controller~\cite{li2015overview}}
        \label{fig:comp}
    \end{figure}

Platoons and their formation and components have been studied extensively \cite{li2017platoon, li2015overview, heinovski2018platoon, deng2023cooperative, li2017dynamical}. The key components of a platoon of vehicles are given in Figure~\ref{fig:comp}\mbox{\cite{li2015overview}} and their description is provided as follows~\mbox{\cite{li2015overview}}:
\subsection{Node Dynamics}
The node dynamics denote the longitudinal dynamics of the vehicles. 
The vehicle longitudinal dynamics are inherently nonlinear, which is composed of engine, transmission, drive line, brake system,
aerodynamics drag, tire friction, rolling resistance,
gravitational force, etc. The commonly used models include: 1)
single integrator model, 2) second-order model (including
double-integrator model), 3) third-order model, and 4)
single-input-single-out (SISO) model~\cite{li2015overview}.   

\subsection{Information Flow Topologies (IFTs)} 
\label{IFT}
Information Flow Topologies (IFTs)~\cite{zheng2016stability} refers to the structure of how the vehicles exchange information with other vehicles in the platoon making them a key factor in fuel efficiency. In fact, communication structures—such as those allowing each vehicle to access data from multiple predecessors—enable smoother and more timely responses to speed changes~\cite{li2025influence}. In Section~\ref{IFTs}, we highlight the importance of IFTs through selected findings from the literature.
\begin{figure}[htb!]
        \centering
        \includegraphics[width=0.7\linewidth]{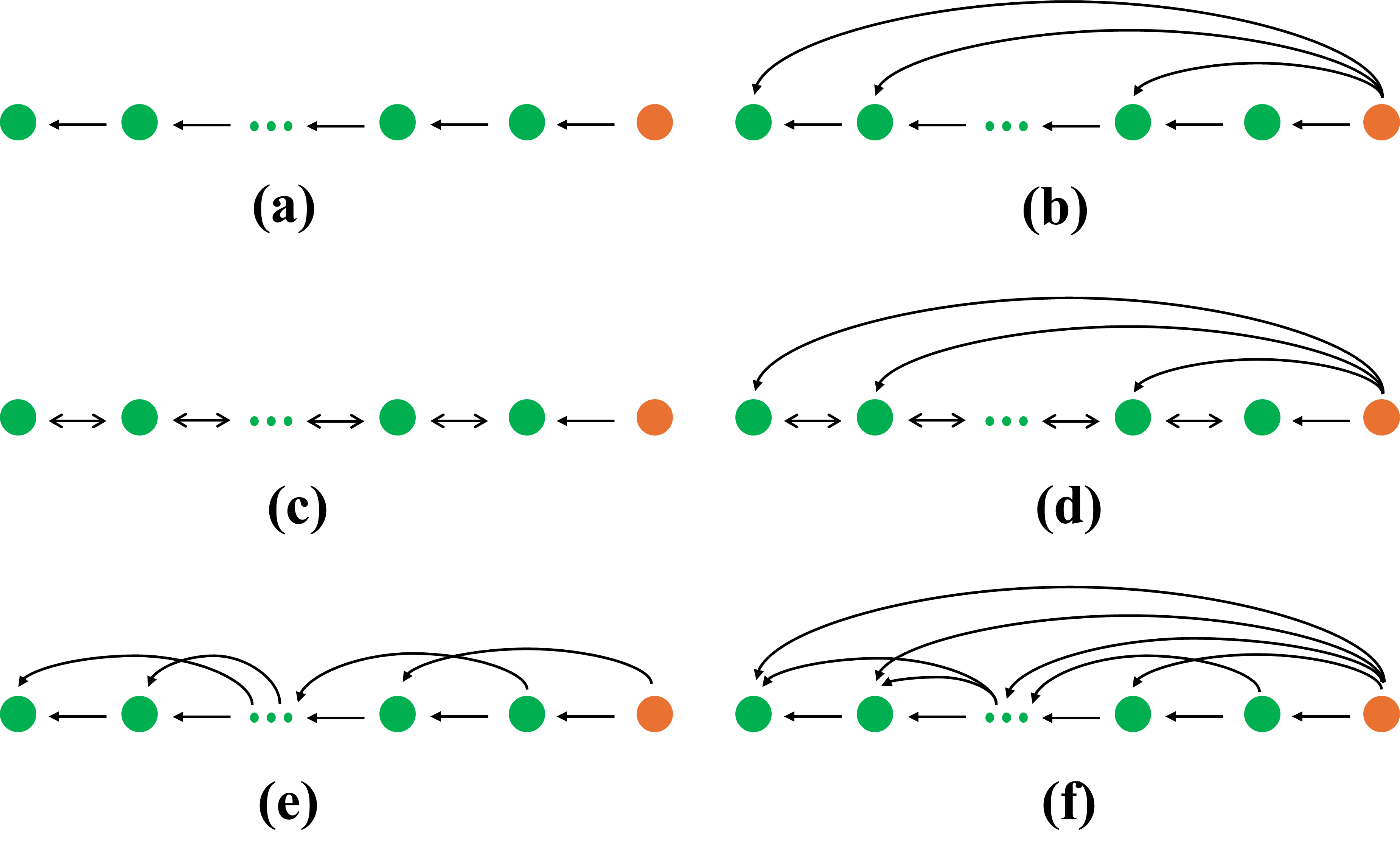} 
        \caption{Typical IFTs for Platoons (a) PF; (b) PLF; (c) BD; (d) BDL; (e) TPF; (f) TPLF~\cite{li2015overview}}
        \label{fig:IFT}
\end{figure}
The commonly used topologies are described below and depicted by Figure~\ref{fig:IFT}, where several communication topologies are used in vehicle platooning systems. For instance, in Predecessor Following (PF)\cite{Sheikholeslam1993Longitudinal}, each vehicle follows its immediate predecessor by adjusting its behavior based solely on the predecessor’s movements and signals. Predecessor Leader Following (PLF)\mbox{\cite{Guo2011}} extends this approach by incorporating information from both the preceding vehicle and the platoon leader to maintain constant intra-platoon spacing. Bidirectional (BD) topology~\mbox{\cite{Zheng2014influence}} further enhances control by considering data from both the preceding and following vehicles, such as relative distance and velocity. In Bidirectional Leader (BDL) topology~\mbox{\cite{Lestas2007}}, the leader communicates with all follower vehicles, and feedback from each follower is sent back to the leader. Two-Predecessor Following (TPF)\mbox{\cite{Swaroop1999constant}} improves stability by allowing each vehicle to monitor both its immediate predecessor and the vehicle ahead of that one. Finally, Two-Predecessor-Leader Following (TPLF)\mbox{\cite{Ploeg2014controller}} enables each vehicle to use information from the immediate predecessor, a second vehicle further ahead, and the platoon leader to make more informed control decisions.

\subsection{Formation Geometry (FG)} Formation Geometry refers to the spatial arrangement of vehicles within a platoon by specifying the desired inter-vehicle distance maintaining the safety, efficiency, stability, and coordination among the vehicles within the platoon. In the following, we provide a brief overview of commonly used spacing policies along with their mathematical formulations. Section~\ref{Enroute} then reviews numerous selected studies from the literature that demonstrate the effectiveness of en-route platoon coordination in optimizing fuel consumption.
\vspace{0.2cm}
\paragraph*{\textbf{Spacing Policies}}
\label{SP}
A platoon formation consists of a string of vehicles traveling as a flock with a tight inter-vehicle distance. This distance is determined based on a chosen spacing policy. In general, there exists two types of intra-platoon spacing policies~\cite{Swaroop1999constant}: \textit{constant spacing}; where it maintains a fixed value of the distance between vehicles independently from their speed, and \textit{variable spacing}~\cite{seiler2004disturbance} that adjusts the distance between vehicles based on their speed. As a representation of both policies, the \textit{constant time-headway spacing} ($CTH$) policy is a popular choice for modern platooning systems representing the distance between vehicles, that is dynamically adjusted based on their speed to maintain a consistent time gap. This time gap is known as the \textit{time headway} ($T_H$) and defined as the time between two consecutive vehicles traveling on the same direction and passing the same point~\cite{tordeux2010adaptive}. Mathematically, $T_H$ is defined as follows~\mbox{\cite{Swaroop1999constant}}:
\begin{equation}
    T_H = \frac{x_l(t) - x_f(t) - L}{v_f}
    \label{eqTime}
\end{equation}where $x_l(t)$ and $x_f(t)$ represent the position of the leader and following vehicles, respectively. \textit{L} is the length of the leading vehicle and $v_f(t)$ refers to the speed of the following vehicle. In this context, the spacing error between vehicles is defined as follows~\mbox{\cite{seiler2004disturbance}}:
\begin{equation}
    e_i(t) = x_{i-1}(t) - x_i(t) - \delta
    \label{eq24}
\end{equation}where $\delta$ is a positive constant representing the desired spacing.

\section{Factors Effective in Reducing Fuel Consumption in Platoons}
\label{sec2}
Reducing energy consumption in vehicle platoons depends on a combination of operational strategies, vehicle technologies, and control mechanisms. In this section, we highlight the key factors that contribute to improved energy efficiency in platooning systems.

\subsection{Aerodynamics Benefits}
\begin{figure}[htb!]
        \centering
        \includegraphics[width=0.7\linewidth]{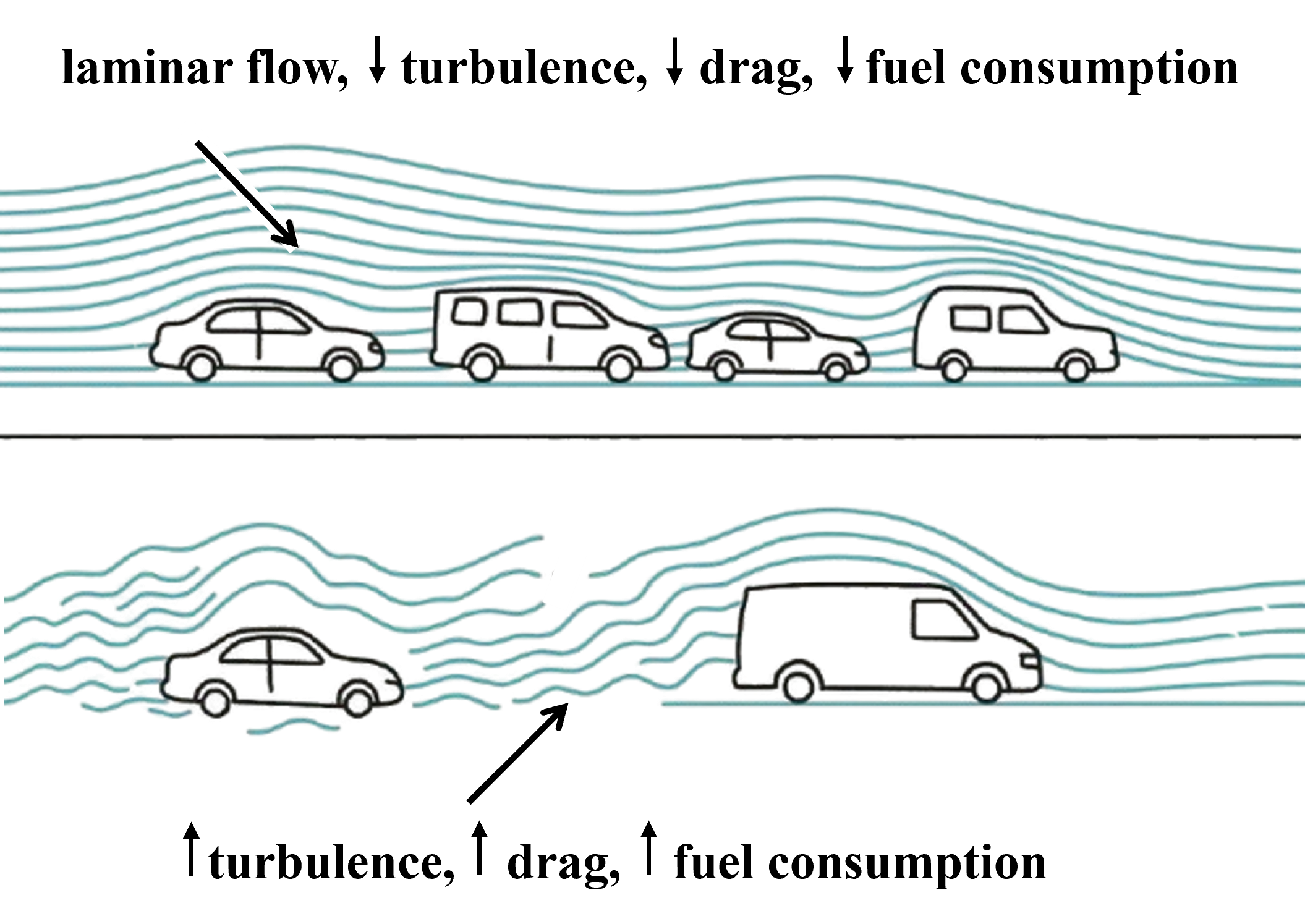} 
        \caption{Illustration of Aerodynamic Effects in Platooning}
        \label{fig:aero}
    \end{figure}
Aerodynamic drag refers to the force opposing a vehicle's motion through the air, resulting from pressure differences and frictional forces along the vehicle’s surface. As a vehicle moves forward, it displaces the air in front of it, creating a high-pressure zone at the front and a low-pressure wake behind. The resulting pressure differential, along with viscous shear forces, generates drag. Thus, when a car follows closely behind another, it experiences significantly less air resistance due to reduced drag \cite{pi2023automotive}. Evidently, one of the main reasons platoons contribute to reduced energy consumption is the aerodynamic advantage they offer to the following cars in a platoon, as depicted by Figure~\ref{fig:aero}. Also, the leading car benefits from the stabilization of air flow behind it. In this section, we review research efforts that model and simulate the aerodynamic effects within vehicle platoons and explain how these effects translate into improved fuel efficiency.
\vspace{0.2cm}
\subsubsection{\textbf{Governing Equations}}
The aerodynamic forces acting on vehicles moving at highway speeds are governed by the principles of fluid dynamics, particularly the behavior of incompressible, turbulent airflow. As vehicles move through the air, they interact with the surrounding airflow, generating aerodynamic forces, e.g., primarily drag and lift. In the context of platooning, understanding and modeling these aerodynamic effects are crucial for optimizing vehicle spacing, reducing drag, and achieving substantial improvements in fuel efficiency.
The governing equations for steady, incompressible, and turbulent flow fields are given as follows~\cite{jo2022numerical}:
\begin{itemize}
    \item \textbf{Continuity equation:}\\

\begin{equation}
\frac{\partial U_i}{\partial x_i} + \frac{\partial U_j}{\partial y_i} + \frac{\partial U_k}{\partial z_i} = 0
\end{equation}

\item \textbf{Momentum equation:}\\
\begin{equation}
    \begin{aligned}
        \frac{\partial U_i}{\partial t} &+ \frac{\partial}{\partial x_j}(U_i U_j) 
        = -\frac{1}{\rho} \frac{\partial P}{\partial x_i} \\
        &\quad + \frac{\partial}{\partial x_j} \left[ \nu \left( \frac{\partial U_i}{\partial x_j} + \frac{\partial U_j}{\partial x_i} \right) - \overline{u_i' u_j'} \right] \\
        &\quad - g_i 
    \end{aligned}
\end{equation}

\end{itemize}
where the Reynolds stress term $\overline{u_i' u_j'}$ is modeled using the Boussinesq approximation:

\begin{equation}
-\overline{u_i' u_j'} = \nu_t \left( \frac{\partial U_i}{\partial x_j} + \frac{\partial U_j}{\partial x_i} \right) - \frac{2}{3}k \delta_{ij}.
\end{equation}where \( U_i \) are the mean velocity components, \( P \) is the mean pressure, \( \rho \) is the fluid density, \( \nu \) is the kinematic viscosity, \( \nu_t \) is the turbulent viscosity, \( k \) is the turbulent kinetic energy, \( \delta_{ij} \) is the Kronecker delta, and \( g_i \) are the gravitational acceleration components.

The $\kappa$–$\epsilon$ turbulence model (KECHEN) is commonly used to simulate turbulent flows~\cite{chen1987computation}. It consists of the following equations\cite{chen1987computation}:
\vspace{0.2cm}
\begin{itemize}
    \item \textbf{Turbulent kinetic energy equation}:
    \vspace{0.2cm}
    \begin{equation}
    \frac{\partial}{\partial x_i}(U_j k) = \frac{\partial}{\partial x_i} \left[ \left( \nu + \frac{\nu_t}{\sigma_k} \right) \frac{\partial k}{\partial x_j} \right] + G - \epsilon
    \end{equation}
    where 
    \[
    G = -\overline{u_i' u_j'} \frac{\partial U_i}{\partial x_j}, \quad \nu_t = C_\mu \frac{k^2}{\epsilon}
    \]
    
    \item \textbf{Energy dissipation rate equation}:
    \begin{equation}
    \frac{\partial}{\partial x_i}(U_j \epsilon) = \frac{\partial}{\partial x_i} \left[ \left( \nu + \frac{\nu_t}{\sigma_\epsilon} \right) \frac{\partial \epsilon}{\partial x_j} \right] + \frac{\epsilon}{k} (C_{\epsilon 1} G - C_{\epsilon 2} \epsilon)
    \end{equation}
\end{itemize}where \( \nu \) is the kinematic viscosity, \( \nu_t \) is the turbulent viscosity, \( \sigma_k \) and \( \sigma_\epsilon \) are the turbulent Prandtl numbers for \( k \) and \( \epsilon \), respectively, and \( G \) is the production of turbulent kinetic energy.
\vspace{0.2cm}
\subsubsection{\textbf{Aerodynamic Pressure Drag}}
The drag force ($F_D$) and drag coefficient ($C_D$) of the model vehicle are calculated using the following equations~\cite{jo2022numerical}:
 \vspace{0.2cm}
\begin{itemize}
    \item \textbf{Drag force} ($F_D$):
    \begin{equation}
        F_D = C_D \cdot \frac{1}{2} \rho_{\text{air}} A_y V_{\text{HDV}}^2
    \end{equation}where \( C_D \) is the drag coefficient, \( \rho_{\text{air}} \) is the air density, \( A_y \) is the projected frontal area of the vehicle in the longitudinal direction, and \( V_{\text{HDV}} \) is the velocity of the heavy-duty vehicle.

    \vspace{0.2cm}
    \item \textbf{Drag coefficient} ($C_D$):
    \begin{equation}
        C_D = \frac{2 F_D}{\rho_{\text{air}} A_y V_{\text{HDV}}^2}
    \end{equation}
\end{itemize}

\subsubsection{\textbf{Fuel Savings and \texorpdfstring{CO$_2$}{CO2} Reduction}}
To assess the impact of platooning, key metrics like traction power, fuel consumption, and CO\textsubscript{2} emissions are used. These metrics quantify efficiency gains from reduced aerodynamic drag and support data-driven evaluation of platooning benefits.
\begin{itemize}
    \item \textbf{Traction Power Saved on Each Model Vehicle Platooning:}\\
The reduction in the tractive power of each model vehicle during platooning, compared to a single moving vehicle (SV), can be calculated as~\ref{eq:pw_sav}:
\begin{equation}
\text{PW}_{\text{sav}} = (F_{D,\text{single}} - F_{D,\text{platoon}}) \times V_{\text{HDV}}
\label{eq:pw_sav}
\end{equation}where \( \text{PW}_{\text{sav}} \) is the tractive power saved [kW], \( F_{D,\text{Single}} \) is the drag force of a single moving vehicle (SV) [kN], \( F_{D,\text{platoon}} \) is the drag force experienced by each vehicle in the platoon (e.g., FV1, FV2, FV3) [kN], and \( V_{\text{HDV}} \) is the velocity of the model vehicle [m/s].

\vspace{0.2cm}
    \item \textbf{Fuel Saved on Each Model Vehicle Platooning Compared to SV:}\\
The mass flow rate of the fuel saved due to aerodynamic drag reduction can be calculated from the power saved using the following equations~\cite{jo2022numerical}:   

\begin{equation}
\dot{m}_{\text{fuel}} = \frac{\text{PW}_{\text{sav}}}{Q_{\text{LHV}} \times \eta_{\text{engine}}}
\label{eq:mfuel}
\end{equation}

\begin{equation}
\dot{Q}_{\text{fuel}} = \frac{\dot{m}_{\text{fuel}}}{\rho_{\text{fuel}}}
\label{eq:Qfuel}
\end{equation}where \( \dot{m}_{\text{fuel}} \) is the fuel mass flow rate [kg/s], \( \text{PW}_{\text{sav}} \) is the tractive power saved [kW], \( Q_{\text{LHV}} \) is the lower heating value of diesel fuel [kJ/kg], \( \eta_{\text{engine}} \) is the brake thermal efficiency of the diesel engine, and \( \rho_{\text{fuel}} \) is the density of diesel fuel [kg/L].

The fuel consumption ($fc$) of a vehicle due to aerodynamic resistance is then given by:
\begin{equation}
fc = \frac{V_{\text{HDV}}}{\dot{Q}_{\text{fuel}}} \quad (\text{km/liter})
\label{eq:fc}
\end{equation}
\noindent where $V_{\text{HDV}}$ is the velocity of the vehicle [km/h].
\vspace{0.2cm}
\item \textbf{Reduction in Carbon Dioxide:}\\
The reduction in CO$_2$ emissions is calculated by this equation~\cite{jo2022numerical}:
\[
\text{CO}_2 \, \text{emissions} = \sum_{i} (AL_i \cdot CL_i \cdot OF_i) \cdot \frac{44}{12}
\]
where \( \text{CO}_2 \) emissions refer to the incineration of fossil liquid waste, \( AL_i \) is the amount of incinerated fossil liquid waste type \( i \), \( CL_i \) is the carbon content of fossil liquid waste type \( i \) (fraction), \( OF_i \) is the oxidation factor for fossil liquid waste type \( i \) (fraction), and \( \frac{44}{12} \) is the conversion factor from carbon to \( \text{CO}_2 \).

\end{itemize}

\subsection{Experimental Studies on the Effect of Drag Reduction in Platoons}
\label{AED}
One of the first successful efforts to measure drag reduction in vehicle platoons was carried out as part of the California PATH project at UC Berkeley and the University of Southern California~\cite{zabat1995aerodynamic}. The study by Zabat \textit{et al.}~\cite{zabat1995aerodynamic}, conducted under the California PATH program, examined the aerodynamic performance of vehicle platoons, focusing on how inter-vehicle spacing affects drag. Using 1/8 scale models of a 1991 GM Lumina APV minivan in wind tunnel experiments, the researchers measured drag forces to quantify the aerodynamic benefits of platooning. They found that both leading and trailing vehicles experienced reduced drag when traveling in close formation, with the greatest benefits observed at shorter distances between vehicles. However, the improvements diminished as spacing increased. These findings suggested that platooning could lead to significant fuel savings due to reduced aerodynamic resistance, providing early empirical support for the development of automated and connected vehicle technologies. 

In a more recent study conducted by the National Research Council Canada (NRC), the authors of~\cite{mcauliffe2017fuel} presented the results of a collaborative study evaluating the fuel-saving potential of truck platooning using Cooperative Adaptive Cruise Control (CACC). Tests were conducted under various conditions, revealing that a three-truck platoon can achieve net fuel savings between 5.2\% and 7.8\%, with improvements up to 14.2\% when combined with aerodynamic trailer devices at short separation distances (17.4 meters). Fuel savings were greatest for trailing vehicles, while the lead vehicle saw minimal benefit. Shorter separation distances led to more significant savings, but gains plateaued beyond about 22 meters for standard trailers and 34 meters for aerodynamic ones. Trucks with aerodynamic trailers consistently outperformed those with standard trailers, and empty trailers achieved slightly better savings than loaded ones. These findings support the potential of platooning to enhance freight transport efficiency and inform future deployment strategies.

For instance, in~\cite{jo2022numerical}, the authors rely on these equations to analyze the aerodynamic interaction in a platoon of four heavy-duty vehicles (HDVs) with the change in the platooning conditions on a freeway. This study highlights significant aerodynamic benefits due to to platooning, where the drag force was reduced by 51\%, 56\%, and 52\% for the platooning vehicles. This implies lower engine power requirements, contributing to improved fuel efficiency and CO$_2$ emission reduction. However, this study only considered steady driving conditions on a level road without side winds, and the aerodynamic characteristics were theoretically analyzed to estimate vehicle energy savings. Furthermore, more research is needed to investigate real-world driving conditions.

In another efforts to examine the impact of platooning on aerodynamic drag reduction, the authors of~\cite{vohra2018examination} investigated vehicle spacing as a key parameter to achieve platoon efficiency. To this end, they conducted a series of experiments to estimate the drag coefficient of 2- and 3-member platoon combinations of medium-duty Volvo VNL-300 day-cab trucks. Subsequently, the relative drag of the leading, middle and trailing trucks are compared by recording the axial drag forces on 1/64 scale models in a wind tunnel. The findings of this study indicate that drag reductions are approximately 36\% at inter-vehicle spacing distances of $3m$ for full-scale 2-truck platoons.

Unlike the studies mentioned above, the work of~\cite{davila2013environmental} conducted real-world fuel consumption tests involving two trucks and three cars in various platoon configurations, where different leading vehicles were used to study the influence of vehicle geometry on fuel consumption, and various inter-vehicle gaps were tested to evaluate aerodynamic effects. By relying on the governing equations, CFD (Computational Fluid Dynamics) simulations were used to assess aerodynamic benefits, particularly reductions in drag coefficient (Cx). CFD simulations demonstrated significant reductions in the drag coefficient, Cx, for following vehicles, with noticeable benefits even for the lead vehicle. Estimated fuel savings ranged from 7\% to 15\% at an 8-meter gap, decreasing to 2\% to 11\% at a 15-meter gap. These fuel savings are directly associated with reductions in CO$_2$ emissions, with environmental benefits measurable in tons of CO$_2$ saved annually.

In a similar setting, the work of~\cite{kaluva2020aerodynamic} investigated the influence of inter-vehicle distance, platoon size and vehicle speed on the drag coefficient of the vehicles in a platoon using CFD. However, the main focus was on analyzing the energy consumption of electric vehicle platoons. To do so, the authors considered two vehicle models, i.e., a minibus and a passenger car, to characterize the drag coefficients of the respective platoons, and conduct an energy analysis to evaluate energy savings. The results of this study, consistent with other findings in the literature, demonstrated a reduction in the average drag coefficient of the platoon of up to 24\% at an inter-vehicle distance of $1 m$ depending on the number of vehicles in the platoon. With a larger inter-vehicle distance of $4 m$, the reduction in the drag coefficient decreased to 4\% of the drag coefficient of the isolated vehicle. Based on these aerodynamic improvements, the study estimated potential energy savings of up to 10\%, depending on factors such as the driving cycle, inter-vehicle distance, and platoon size.

\subsection{En-Route Platoon Coordination in Minimizing Fuel Consumption}
\label{Enroute}
En-route platoon coordination is a real-time strategy for forming and managing groups of vehicles in the form of platoons during a trip to minimize overall fuel consumption. Unlike pre-scheduled or static platooning, it dynamically identifies opportunities for vehicles traveling in the same direction to merge into or leave platoons based on traffic conditions, vehicle routes, and energy efficiency goals. For instance, the study by~\cite{hall2005vehicle} focused on strategies for forming vehicle platoons in Automated Highway Systems (AHS) to enhance roadway throughput, fuel consumption, and safety. To achieve this, the paper explored sorting vehicles by destination at highway entrance ramps, aiming to align vehicles with similar routes into the same platoons. Various platoon formation strategies were evaluated based on key metrics: platoon size, throughput, and formation time. The study found that dedicated assignments based on destination improved platoon size and throughput, though they resulted in slightly longer waiting times at entrances.

In a more in-depth study of the impact of en-route formation on fuel consumption, This study focuses on maximizing fuel savings for heavy-duty vehicles traveling in platoons by introducing a distributed control approach across a road network. The authors propose local controllers that use minimal information, such as vehicle position, speed, and destination, to make quick speed adjustment decisions aimed at enabling future platoon formations. For small vehicle sets, the optimal control and routing problem is solved exactly, while a fast heuristic is developed for large-scale, real-time implementation. The method is validated through a large-scale simulation of the German autobahn, involving thousands of vehicles. Results show fuel savings ranging from 1–9\%, with gains above 5\% when only a small fraction of vehicles participate. While the study generally assumes vehicles do not travel longer than their shortest path travel time, the authors note that relaxing this assumption can lead to even greater fuel savings.

Building on the idea of distributed control for large-scale platoon coordination, the study by Van \textit{et al.}~\cite{van2015fuel} addressed the more centralized challenge of coordinating platoon formation and breakup in a fuel-optimal manner. The authors formulated an optimization problem that considered routing, speed-dependent fuel consumption, and platooning decisions. An approximate solution was proposed, starting with the shortest path computation for each truck, followed by identifying potential platoon configurations. When tested in a realistic scenario, the method demonstrated promising results, effectively coordinating platoon formation and speed profiles to minimize fuel consumption while respecting operational constraints.

Moreover, Liang \textit{et al.}~\cite{liang2015heavy} explored how two or more scattered vehicles can cooperate to form platoons efficiently. They demonstrated that when platoons are formed on the fly along the same route without considering rerouting, road topography has little impact on the coordination decision. The study proposed an optimization problem for coordinating two vehicles to form a platoon and proposes an algorithm for forming platoons by coordinating neighboring vehicles pairwise. A simulation using detailed vehicle models and real road topography showed that this approach leads to significant fuel savings. 

In contrast to the localized study in~\cite{liang2015heavy} focusing on fuel-efficient en-route platoon formation without considering rerouting or road topography impact, the study by Van \textit{et al.}~\cite{van2017fuel} scaled the concept to large-scale fleet coordination that involves many trucks with differing routes, delivery deadlines and includes rerouting options to maximize platooning opportunities across the network. In this study, the authors investigated strategies for coordinating heavy-duty vehicle platooning during trips to maximize fuel efficiency. By modeling the platoon formation as a combinatorial optimization problem and applying heuristic and convex optimization methods, they demonstrated that significant fuel savings can be achieved even in large-scale scenarios. Their simulations confirmed that dynamic, en-route platoon coordination is both practical and highly effective for reducing overall fuel consumption across a fleet.\\
\indent In another study more focused on route vehicle management~\cite{cerutti2021aerodynamic}, the authors conducted an experimental study to investigate the aerodynamic behavior of light commercial vehicles traveling in a platoon configuration, focusing on drag measurements and pressure distribution. Results showed that drag reduction is most effective at small inter-vehicle distances (especially between 0.5 and 1 vehicle lengths), while the benefit significantly decreases for distances greater than three vehicle lengths (d/L $>$ 3). The study reveals that reducing spacing affects the wake structure behind the leading vehicle, notably suppressing a dominant vortex (V2), which plays a key role in the flow dynamics and drag behavior. The findings support previous numerical studies and highlight the role of flow control, vortex dynamics, and vehicle interaction in reducing drag. \\
\indent Furthermore, several studies have explored en-route platoon coordination to minimize fuel consumption, demonstrating consistent benefits through collaborative vehicle grouping. Zhen \textit{et al.}~\cite{zeng2022decentralized} and Liu \textit{et al.}~\cite{liu2024decentralized} proposed a decentralized, traffic-adaptive approach for real-time platoon formation, demonstrating significant fuel savings by optimizing merge points under varying traffic conditions. The authors of~\cite{ard2022simulated} investigated fuel-efficient platooning via model predictive control (MPC), achieving a 9\% reduction in consumption under varying traffic conditions, while in~\cite{rezgui2020platooning}, the authors integrated V2V communication for adaptive platoon reconfiguration, further lowering energy use. Similarly,~\cite{dokur2023v2v} introduced a hybrid V2I/V2V-based coordination strategy for highway merging, achieving 12–18\% fuel reduction through enhanced communication between vehicles and infrastructure. Yang \textit{et al.}~\cite{yang2024efficient} employed a rolling horizon control method to dynamically assemble platoons at on-ramps, improving fuel economy by 8–14\% compared to uncoordinated merging. Finally, Wang \textit{et al.}~\cite{wang2024deep} leveraged reinforcement learning to optimize platoon coordination in mixed-autonomy traffic, reducing fuel consumption by 10–15\% while adapting to real-time driving behaviors. Collectively, these studies underscore the effectiveness of adaptive control strategies, V2X communication, and AI-driven optimization in minimizing fuel use through intelligent en-route platoon coordination.

\subsection{Speed Optimization within Platoons Help with Fuel Optimization}
Speed optimization within platoons is a control-based approach that adjusts the velocity profiles of platooning vehicles to minimize overall fuel consumption. By coordinating speeds based on vehicle dynamics, traffic conditions, and road topology, this method ensures smoother driving with fewer accelerations and decelerations~\cite{horowitz2002control, marcano2020review, rajamani2000design, jin2018experimental}. Often integrated with platoon planning algorithms, it enhances fuel efficiency beyond the aerodynamic benefits of close-following alone, making it a key component in intelligent fuel-saving strategies for connected and automated vehicles.

Several studies have explored the potential of platooning and speed coordination to reduce fuel consumption and emissions. A traditional approach proposed by~\cite{caltagirone2015truck} represented and compared various truck platooning strategies across different road topographies. Using a stochastic optimization of lead vehicle speed profiles (LV-SPO) over 10 km segments, it achieved fuel savings of about 15.4\% compared to standard cruise control. For the following vehicles, models based on artificial physics, particularly one using modified artificial gravity, were tested to minimize total platoon fuel consumption. This gravity-based model slightly outperformed standard adaptive cruise control while preserving safety and platoon stability.

While LV-SPO focuses solely on optimizing the Speed Profile of the Lead Vehicle, it may not fully account for the dynamics between all vehicles in the platoon. This limitation can reduce overall fuel efficiency. To address this, the platooning Speed Profile Optimization (P-SPO) method proposed by~\cite{torabi2018fuel}, which focused on optimizing individual speed profiles for each heavy-duty vehicle (HDV) within a platoon. Evaluated in simulation over 10 km road segments, P-SPO demonstrated fuel savings of 15.8\% for homogeneous platoons and up to 17.4\% for heterogeneous ones, outperforming standard cruise control and adaptive cruise control setups.

In another study, Guo \textit{et al.}~\cite{guo2018fuel} introduced a two-layer hierarchical framework for truck platoon coordination, focusing on speed planning and tracking control. Unlike the previous methods such as LV-SPO and S-SPO, which either optimize only the lead vehicle's speed profile or optimize all vehicles' speed profiles in a uniform manner, this approach uses an average vehicle to calculate the speed profile. This helps improve fuel efficiency, especially for heterogeneous platoons with varying vehicle weights and sizes. Additionally, the control layer employs a discrete-time back-stepping control law, considering factors like road slope and vehicle heterogeneity, with a novel string stability criterion to enhance platoon stability. 

Moreover, \cite{alam2013look} introduced a Look-ahead Cruise Control that was able to reduce fuel consumption up to 14\% over the steep slopes of the road. Building on this concept, several studies have explored cooperative look-ahead control strategies aimed at maximizing fuel efficiency within truck platoons. For instance, \cite{turri2014fuel} proposed a distributed control framework that allowed each vehicle in the platoon to utilize preview information about road grade and traffic conditions to optimize its velocity profile while maintaining safe inter-vehicle distances. Similarly, \cite{johansson2017look} focused on dynamic programming-based optimal speed planning for heavy-duty vehicle platoons, particularly in deceleration scenarios triggered by detected traffic slowdowns ahead. Their approach accounted for maximum acceleration constraints and demonstrated improved fuel economy and travel time efficiency. In contrast, \cite{turri2018fuel} investigated ad-hoc, non-cooperative platooning, where a following vehicle optimizes its speed by using road slope data and the preceding vehicle’s speed trajectory. Their dynamic programming-based look-ahead adaptive cruise controller was able to achieve fuel savings of up to 7\% by adjusting inter-vehicular distances in response to upcoming slopes. The authors of~\cite{hu2021eco} introduced an ecological and string-stable control scheme (E-CACC), defining novel spacing policies and control laws to ensure platoon members followed energy-optimal trajectories, improving fuel efficiency while maintaining safe and stable platoon behavior. Lastly, \cite{zhai2018cooperative} formulated a cooperative fuel-optimal control problem using distributed MPC that accounts for aerodynamic drag, fuel consumption models, and discrete gear ratios, improving fuel efficiency under various conditions.

Other studies have also shown that reducing the distance between vehicles in a platoon can improve fuel efficiency. For instance, a CFD study, in~\cite{humphreys2016computational}, was conducted to examine how inter-vehicle distances affect fuel consumption in a Driver-Assistive Truck Platooning (DATP) system. To validate their approach, a series of simulations was performed at multiple separation distances (30 ft, 40 ft, 50 ft, 75 ft, and 150 ft). These tests aimed to correlate drag reduction trends with spacing showing that fuel economy improves as separation distance decreases. In~\cite{sidorenko2020vehicle}, the authors introduced protocols that adjust the distance between trucks in a platoon based on the quality of V2V communication, particularly in emergency braking situations. By temporarily reducing braking power, the platoon can maintain closer distances between trucks, which enhances fuel economy. Similarly, in~\cite{tornell2021influence}, the authors examined the aerodynamics of closely spaced vehicles using CFD. The study showed that as the inter-vehicle distance decreased from 20 m to 2.5 m, the combined drag of the platoon continuously improved improving fuel efficiency.

\subsection{The Impact of Platoon Information Flow Topologies on Fuel Consumption}
\label{IFTs}
One of the key factors influencing the efficiency of a vehicle platoon is the information flow topology (IFT)~\cite{Zhang1999using}, the structure through which information such as speed, acceleration, and braking commands is communicated among vehicles. An overview of common IFTs is presented in Section~\ref{IFT}, with Figure~\ref{fig:IFT} illustrating the most widely studied architectures in the literature. 

Since the present review focuses on fuel consumption in the context of platooning, it is important to note that fuel savings primarily stem from two sources: 1) reduced aerodynamic drag due to close inter-vehicle spacing, and 2) smoother driving behavior enabled by better coordination of acceleration and braking. While the first aspect is analyzed in detail in Section~\ref{AED}, the second is directly influenced by the choice of IFT. Specifically, topologies that allow for earlier and more accurate responses to the lead vehicle's behavior help reduce unnecessary acceleration and braking events, thereby improving energy efficiency and lowering overall fuel consumption.

Yan \textit{et al.}~\cite{yan2022pareto} proposed a novel method for determining Pareto-optimal IFTs in vehicle platoons using the NSGA-II algorithm~\cite{deb2002fast}, which was performed offline to enhance overall platoon performance. The impact of various IFTs was assessed across five key criteria: tracking ability, velocity smoothness, fuel economy, resilience to communication delays, and communication efficiency. Key findings show that increasing the number of communication links improves tracking accuracy, while metrics such as velocity smoothness, fuel efficiency, and communication efficiency tend to be aligned but sometimes conflict with tracking performance. The proposed approach is validated through five case studies involving heterogeneous vehicle dynamics and external disturbances. Results demonstrated significant improvements: tracking ability by 33.67\%–49.35\%, fuel economy by 7.18\%–16.93\%, and reduced acceleration variability by up to 14.9\%.

Another study by Zhou \textit{et al.}~\cite{zhou2021impact} has showed that CAV platoons can reduce fuel consumption and emissions by up to 34.7\% and 41.4\%, respectively. However, degradation in IFT —caused by control system fallback— leads to capacity breakdowns and significantly increases fuel use (up to 44.5\%) and emissions (up to 60.1\%). In another effort to highlight the importance of IFTs on fuel consumption, Dang \textit{et al.}~\cite{dong2025analyzing} investigated how the spatial formation of CAVs and HDVs within a platoon affects energy and traffic efficiency. Using heterogeneous car-following models, the authors simulated multiple platoon configurations across different driving modes and CAV penetration rates. Results showed that platoon formation significantly impacts performance, with energy efficiency varying by 1.24\% to 26.27\% and traffic efficiency by 0.49\% to 29.47\% depending on vehicle arrangement.

V2V communication enables advanced platoon control strategies that can enhance transportation efficiency. In this context, Hu \textit{et al.}~\cite{hu2022fuel} proposed a distributed control approach that minimizes fuel consumption by combining switching feedback control with economic model predictive control (EMPC), under a multiple-predecessor following IFT. The MPF topology plays a key role in improving fuel efficiency, as it allows each vehicle to access information from several predecessors, enabling more predictive and coordinated responses. This enhanced situational awareness reduced unnecessary acceleration and braking, which directly improves fuel economy. Simulation results showed that this approach achieved up to 6.84\% fuel savings compared to tracking-based control. 

Table~\ref{tab1} presents a summary of fuel reduction results based on various approaches. It compiles findings from selected studies in the literature, highlighting key factors that influence fuel consumption, including drag reduction, en-route platoon coordination, speed optimization, and information flow topology.

\begin{figure*}[htb!]
        \centering
        \includegraphics[width=0.8\textwidth]{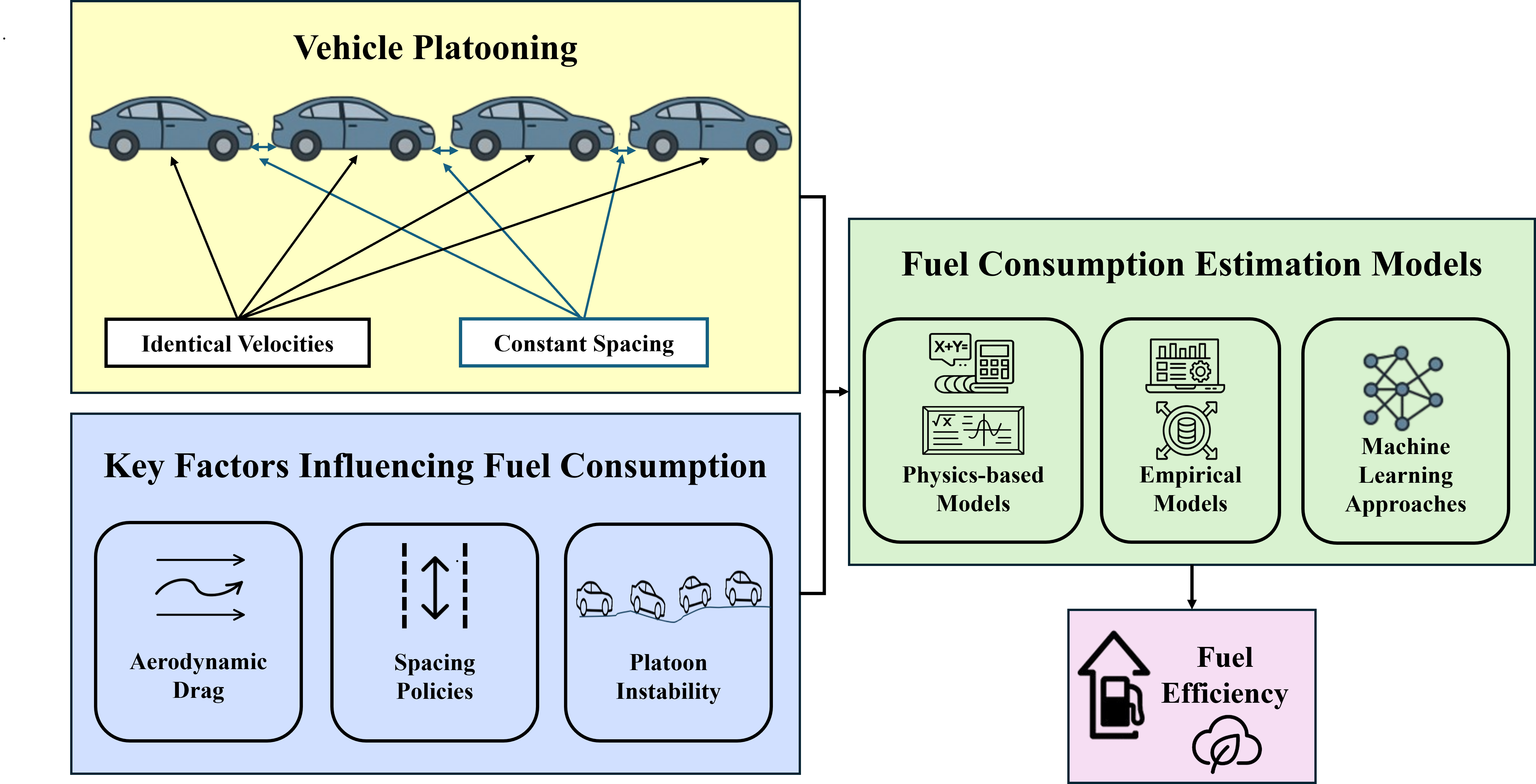} 
        \caption{Key Factors Influencing Fuel Consumption in Vehicle Platooning}
        \label{fig:diag}
\end{figure*}
    
\begin{table*}[htb!]
\caption{Methods Investigated in the Literature for Fuel Consumption Reduction}
\label{tab1}
\renewcommand{\arraystretch}{1.2}
\begin{tabular}{|m{3cm}|p{4.8cm}|p{9cm}|}
\hline
\textbf{Methods for Fuel Reduction} & \textbf{References} & \textbf{Description} \\ \hline

\textbf{Drag Reduction} & 
\cite{steers1977reduced}, \cite{rose1981commercial}, \cite{bonnet2000fuel} &
These experimental approaches involve controlled on-road tests, wind tunnel measurements, and full-scale field trials. These studies focus on evaluating aerodynamic design modifications on drag reduction and fuel efficiency, using empirical data to quantify performance improvements. \\ \cline{2-3}

& \cite{watkins2008effect}, \cite{abinesh2014cfd}, \cite{sivaraj2018reduction} &
These works focus on drag reduction analysis using a combination of wind tunnel experiments and computational simulations. These studies analyze how vehicle geometry, surface features, and inter-vehicle spacing affect airflow and aerodynamic drag, employing tools such as CFD and physical modeling to quantify the potential for drag and fuel consumption reduction. \\ \cline{2-3}

& \cite{patidar2015fuel}, \cite{ali2017design}, \cite{saleh2020numerical} &
These studies focus on numerical studies that use computational modeling and simulations to analyze fuel consumption and aerodynamic performance. These works employ numerical optimization to evaluate vehicle designs, providing insights into potential fuel savings and efficiency improvements. \\ \hline

\textbf{En-Route Platoon Coordination} & 
\cite{fernandes2014multiplatooning}, \cite{hoef2019predictive}, \cite{zhai2020ecological} &
These approaches focus on the design and implementation of distributed control strategies for vehicle platooning. These studies develop and analyze decentralized algorithms that enable individual vehicles to make real-time decisions based on local information, aiming to improve platoon stability, safety, and fuel efficiency through coordinated control without relying on a centralized controller. \\ \cline{2-3}

& \cite{khan2005convoy}, \cite{larson2016coordinated}, \cite{meisen2008data} &
These works focus on developing frameworks for coordinated vehicle convoys. These frameworks integrate communication, control algorithms, and data analytics to enable synchronized movement, improve fuel efficiency, and enhance overall convoy safety and operational performance. \\ \cline{2-3}

& \cite{steinmetz2017fast}, \cite{nourmohammadzadeh2016fuel}, \cite{sokolov2017maximization} &
The approaches focus on routing strategies for vehicle platoons. These studies develop algorithms to optimize routes and coordination among vehicles, aiming to maximize fuel efficiency, reduce travel time, and improve overall traffic flow through intelligent path planning and platoon formation. \\ \hline

\textbf{Speed Optimization} & \cite{miao2018connectivity}, \cite{guo2023optimization}, \cite{xu2018cooperative}, \cite{chang2005vehicle}, \cite{williams2013evaluation}, \cite{Stankovic2000decentralized} & These studies focus on speed analysis aimed at fuel consumption reduction. These studies analyze vehicle speed profiles and coordination strategies to optimize acceleration and deceleration patterns, improving fuel efficiency by minimizing unnecessary speed fluctuations and enhancing platoon smoothness through cooperative control. \\ \hline

\textbf{Information Topology} & \cite{kim2013fuel}, \cite{zheng2015stability}, \cite{li2022topology}, \cite{sorlei2021fuel}, \cite{bian2021fuel}, \cite{zheng2016stability}, \cite{Ghasemi2013stable}, \cite{Wang2023cooperative} & These studies focus on information topology in vehicle platooning. These studies analyze and design communication structures that govern how information is shared among vehicles, aiming to improve platoon stability, control performance, and fuel efficiency by optimizing the flow of data within different network topologies. \\ \hline
\end{tabular}
\end{table*}

\section{Fuel Consumption Estimation in Platoons}
\label{sec3}
Platooning reduces aerodynamic drag, especially for following vehicles, leading to fuel savings and emissions reduction. Quantifying these savings is important for platooning management, policy-making, and intelligent transportation systems. This estimation can be done using physics-based models, empirical models, and machine learning algorithms, as depicted by Figure~\ref{fig:diag}. In this section, we review relevant literature under each of these model categories.

\subsection{Physics-Based Models}
Physics-based models estimate fuel consumption by modeling the vehicle's dynamics and the energy needed to move a vehicle under specific conditions (speed, acceleration, slope, etc.), then converting that energy demand into fuel usage based on engine efficiency \cite{zhou2016review}. A widely referenced physics-based model in the literature is the one proposed by Akcelik \textit{et al.}~\cite{akcelik1989efficiency}, which estimates fuel consumption as a function of vehicle speed and acceleration by capturing key physical effects such as rolling resistance, aerodynamic drag, and the influence of acceleration on engine load. The model distinguishes between positive and negative acceleration, reflecting the increased fuel demand during acceleration phases while ignoring fuel recovery during braking. Its mathematical formulation, providing a relative comparison of fuel consumption between the first and last vehicle in the platoon, is given as follows~\cite{akcelik1989efficiency}:
\begin{equation}
    \begin{aligned}
        F = \max(0, b_1 +b_2v + b_3v^2 + b_4v^3 &\\+ c_1va + c_2v(\max(0,a))^2)
    \end{aligned}
    \label{model}
\end{equation}

\begin{itemize}
    \item $F$: Estimated instantaneous fuel consumption rate (typically in g/s or L/100km, depending on calibration).
    \item $v$: Vehicle speed (in meters per second, m/s).
    \item $a$: Vehicle acceleration (in meters per second squared, m/s²).
    \item $b_1, b_2, b_3, b_4$: Coefficients capturing the \textit{base consumption} and \textit{speed-dependent} components.
    \begin{itemize}
        \item $b_1$: Idle/base consumption (even at zero speed).
        \item $b_2 v$: Linear speed contribution.
        \item $b_3 v^2$, $b_4 v^3$: Non-linear (quadratic/cubic) speed contributions, modeling aerodynamic drag and other effects.
    \end{itemize}
    \item $c_1$: Coefficient for the interaction between speed and acceleration ($v a$), reflecting the increased fuel needed during acceleration.
    \item $c_2$: Coefficient for the term $v(\max(0,a))^2$, which only activates when the vehicle is accelerating (not decelerating), modeling the steep increase in fuel usage during strong accelerations.
\end{itemize}

In the work of Turri \textit{et al.}\cite{turri2016cooperative}, the fuel consumption model is directly tied to the vehicle’s longitudinal dynamics, capturing how various physical forces contribute to fuel demand. The motion of each vehicle is governed by Newton’s second law, incorporating propulsion, braking, gravity, rolling resistance, and aerodynamic drag forces. The aerodynamic drag is especially important in \textit{platooning scenarios}, where the inter-vehicle distance $d_i$ significantly influences the drag coefficient $C_D(d_i)$, and thus, the tractive effort and fuel consumption. The fuel flow $\delta_i$ is modeled as a function of engine speed and power output, with the latter being the product of traction force $F_{e,i}$ and vehicle speed $v_i$. By assuming negligible powertrain losses and simplifying the dependence on gear ratio, the model reduces to a single-variable map:
\begin{equation}
    \delta_i = \phi_{opt,i}(F_{e,i}v_i)
\end{equation}which links the required tractive power directly to the fuel flow. This makes the model both physically interpretable and computationally tractable for optimization and control tasks in platoons.

Another fuel consumption model, proposed in Feng \textit{et al.}~\cite{feng2023controller}, is based on engine data collected from the commercial software TruckSim~\cite{grace1998carnegie}. The model uses a third-order polynomial to represent fuel consumption as a function of engine torque $T_e$ and engine speed $n_e$. The fuel consumption function is given by~\cite{feng2023controller}:
\begin{equation}
\begin{aligned}
f_r(T_e,n_e)=&C_{00}+C_{10}n_e+C_{01}T_e+C_{11}T_en_e+\\&C_{02}n_e^2+C_{20}T_e^2+C_{21}n_e^2T_e+\\&C_{12}T_e^2n_e+C_{03}T_e^3
    \label{fuel_mod}
    \end{aligned}
\end{equation}
where $C_{ii}$ are the polynomial coefficients, $T_e$ is the engine output torque, and $n_e$ represents the engine speed (RPM). This model is a \textit{physics-based engine map}, where the fuel consumption is derived from physical measurements and engine characteristics, rather than from black-box learning methods. 

The engine speed $n_e$ is related to the vehicle’s speed through its gear ratio~\cite{feng2023controller}:
\begin{equation}
    n_e = \frac{i_t}{r} v_x \times \frac{60}{2\pi}
    \label{RPM}
\end{equation}where $i_t$ is the overall gear ratio, $r$ is the wheel effective radius, and $v_x$ is the linear speed (vehicle speed).

The fuel consumption is then used to define a cost function $J_{DP}$ for \textit{dynamic programming}, aimed at minimizing fuel use in platoon scenarios~\cite{feng2023controller}:
\begin{equation}
    J_{DP} = \sum_{i=1}^{\mathcal{N}+1}\sum_{j=k}^{k+N_{DP}} y(j)
    \label{cost}
\end{equation}where $y(j)$ represents the fuel consumption at the $j$-th time step, and $N_{DP}$ is the number of intervals in the dynamic programming strategy. This approach provides an energy-efficient solution for optimal control in cooperative driving, where minimizing fuel consumption is a key objective.

\subsection{Empirical Models}
An empirical model estimates fuel consumption based on experimental or real-world data, typically through regression, lookup tables or curve fitting. These models are built by analyzing measured relationships between variables (e.g., speed, acceleration) and fuel consumption using statistical techniques.

A prominent example is the VT-Micro model introduced by Ahn \textit{et al}.~\cite{ahn2002estimating}, which uses a multivariate polynomial function of instantaneous speed and acceleration to estimate fuel consumption and emissions. To evaluate fuel consumption, the VT-Micro model was used for its simplicity, where its mathematical model is explained as follows~\cite{ahn2002estimating}:
\begin{equation}
    \ln{(MOE_e)} = \sum_{i=0}^{3} \sum_{j=0}^{3} k_{i,j}^e v_n^i(\dot{v}_n)^j
    \label{fuelmodel}
\end{equation}where \( MOE_e \) is the \( n^{\text{th}} \) vehicle's fuel consumption and emission rate, \( i \) and \( j \) are the powers of speed and acceleration, respectively, and \( v_n \) and \( \dot{v}_n \) are the instantaneous speed and acceleration of the \( n^{\text{th}} \) vehicle, respectively. \( k_{i,j}^e \) is a regression coefficient at speed power \textit{i} and acceleration power \textit{j}. This model integrates polynomial regression techniques trained on microscopic driving data, making it particularly suitable for integration into traffic simulation environments. Despite its simplicity and computational efficiency, the model relies purely on statistical fitting, which may limit its physical interpretability and generalization across different vehicle types or driving conditions.

To address these limitations, more physically grounded empirical models have been proposed. One such model is the VeHICLE model (Vehicle-Specific Power model)~\cite{jimenez1998understanding} that is widely used to estimate fuel consumption in various driving conditions. This model is based on Vehicle-Specific Power (VSP) and provides an empirical formula that links fuel consumption to the vehicle's speed, acceleration and road grade. The model is especially useful for urban driving scenarios and has been widely adopted in traffic simulation studies. Fuel consumption is estimated as a function of vehicle speed, acceleration, road grade, and rolling resistance.\\
The VSP-based empirical model is given by~\cite{jimenez1998understanding}:
\begin{equation}
    \delta_i = \alpha_1 v_i + \alpha_2 \dot{v}_i + \alpha_3 g_i
    \label{eq:vehcile_model}
\end{equation}where $v_i$ is the speed of vehicle $i$, $\dot{v}_i$ is its acceleration, and $g_i$ is the road grade at its position.

\subsection{Machine Learning Models}
\begin{figure*}[htb!]
        \centering
        \includegraphics[width=0.5\linewidth]{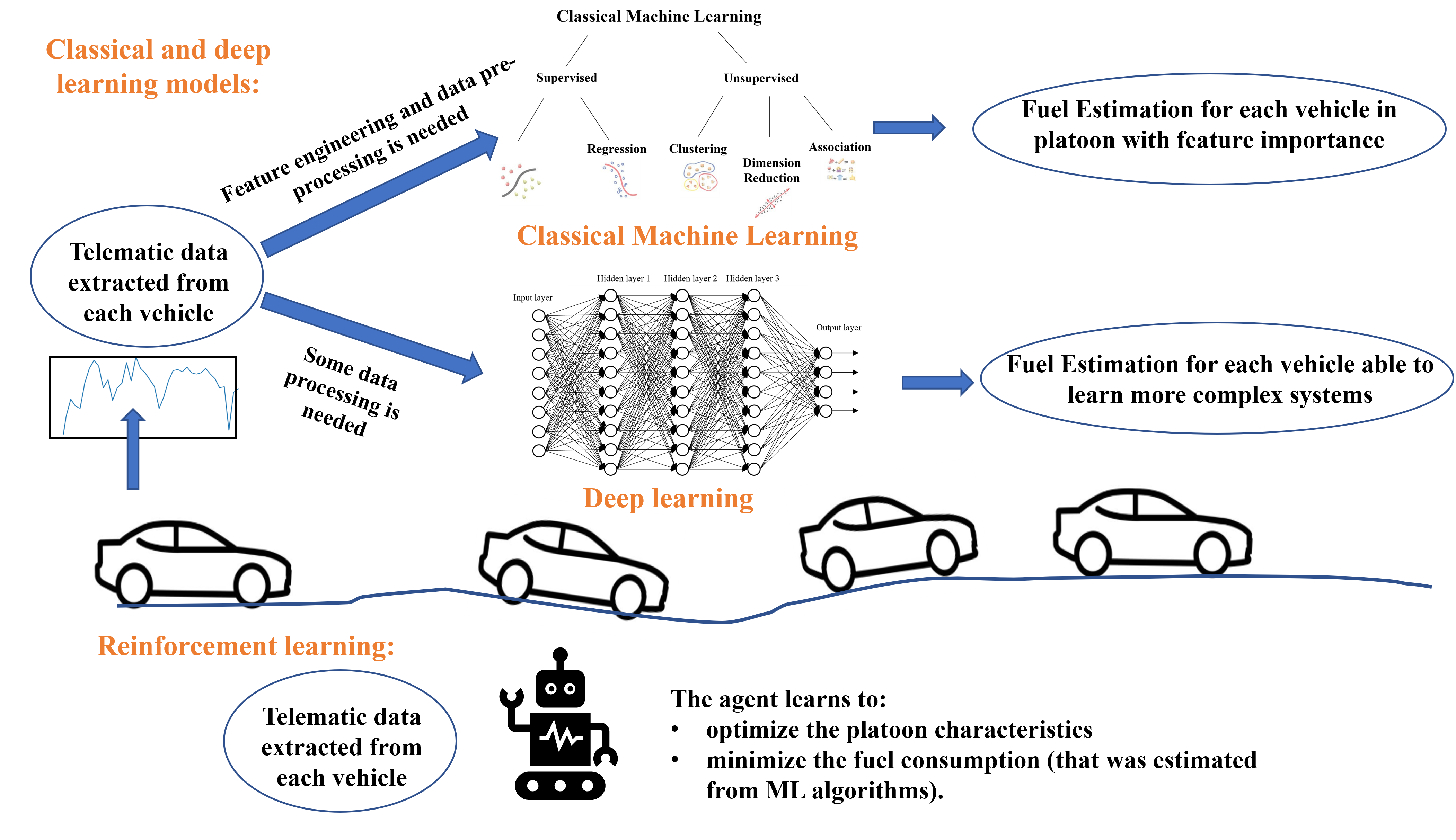} 
        \caption{Illustration of Aerodynamic Effects in Platooning Telematics}
        \label{fig:ML}
\end{figure*}

Machine Learning (ML) models have gained traction recently being employed for fuel consumption estimation due to their ability to learn complex, nonlinear relationships from data without requiring detailed physical modeling. These models are data-driven and often outperform traditional models when sufficient high-quality data is available. Some of the most common ML models for fuel consumption estimation are Support Vector Regression (SVR)~\cite{awad2015svr}, Random Forest (RF)~\cite{rigatti2017random}, Gradient Boosted Trees~\cite{natekin2013gradient}, and neural networks~\cite{abdi1999neural}. Another approach suggested in this context combined hybrid ML techniques with physics models, enabling more accurate and generalizable fuel consumption estimation by leveraging both data-driven insights and physical vehicle dynamics. Although detailed reviews of ML methods for fuel consumption estimation are available in \cite{wickramanayake2016fuel} and \cite{seyedzadeh2018machine}, this section offers a brief summary of the most commonly used algorithms in recent research as depicted by Figure~\ref{fig:ML}.\\

A general formulation of the ML algorithms is provided as follows, where $\mathbf{x}(t)$ denotes the input feature vector at time $t$, which may include variables such as vehicle speed $v(t)$, acceleration $\dot{v}(t)$, road grade $g(t)$, engine torque $T_e(t)$, and other vehicle-specific parameters. The fuel consumption \( \hat{y}(t) \) is estimated using a generic machine learning model expressed as a function of relevant features:
\begin{equation}
    \hat{y}(t) = f_{\theta}(\mathbf{x}(t))
    \label{ml_general}
\end{equation}
where $f_{\theta}$ is a function parameterized by $\theta$ and trained using fuel consumption data.

Support Vector Machines (SVM) and decision tree-based models like Random Forests (RF) and Gradient Boosting Machines (GBM) have proven effective for estimating vehicle fuel consumption. SVMs work by defining margin-based decision boundaries, while tree-based methods rely on information gain to split data and make predictions. These models have been successfully applied using features such as velocity, acceleration, and torque, with Random Forests, for example, achieving high predictive accuracy across several studies~\cite{hassan2023random, breiman2001random, fu2024novel, wen2021features, Luo2023modeling,Liu2024automated}. Support Vector Regression (SVR), a kernel-based variant of SVM, estimates fuel consumption by computing a weighted sum of kernel evaluations between input features and support vectors, allowing it to capture complex, non-linear patterns~\cite{hearst1998support, abukhalil2020fuel,liu2023high,Luo2023modeling}. A key strength of both SVR and decision tree-based models is their ability to provide insights into feature importance, enhancing interpretability and helping identify the most influential factors affecting fuel consumption.

Neural networks are more complex models that can capture nonlinear relationships in data, but do not inherently offer insight into feature relevance. To interpret neural networks, methods such as ablation studies are often used to assess the impact of each input feature. For example, in the case of a feedforward neural network with one hidden layer \cite{gurney2018introduction}:
\begin{equation}
    \hat{y}(t) = \sigma\left( \mathbf{w}_2^T \cdot \phi(\mathbf{W}_1 \mathbf{x}(t) + \mathbf{b}_1) + b_2 \right)
    \label{ml_nn}
\end{equation}where \( \phi(\cdot) \) is the hidden layer activation function (e.g., ReLU or tanh), \( \mathbf{W}_1 \) and \( \mathbf{w}_2 \) are weight matrices, \( \mathbf{b}_1 \) and \( b_2 \) are biases, and \( \sigma \) is the output activation function, typically linear in regression tasks.

Deep learning algorithms built by stacking multiple hidden layers have recently been applied to vehicle fuel estimation, offering improved accuracy by capturing complex, nonlinear relationships in driving data. For example, the authors in~\cite{das2024emissions} proposed a novel car-following model for autonomous vehicles (AVs) using a physics-informed Long Short-Term Memory LSTM Neural Networks~\cite{sak2014long}. The model employs a multi-objective loss function to simultaneously replicate real-world AV trajectories and minimize emissions, integrating Vehicle-Specific Power into its physics-based framework as an emissions proxy. In~\cite{mohamed2024developing}, the authors explored how various truck platoon configurations affect fuel savings using data from 10 field experiments. Four models were compared: vanilla neural network, XGBoost, K-nearest neighbors, and negative binomial regression. The negative binomial model performed best (74\% accuracy) and was used to derive a fuel-savings equation. Key findings showed that larger platoon size and higher speeds yield the greatest fuel savings, conventional trucks outperform cab-over designs, and truck weight has only a minor effect.

In another work~\cite{ling2018fuel}, the authors proposed a Model Predictive Control (MPC) framework to manage the longitudinal dynamics of a follower HDV in a two-truck platoon without relying on vehicle-to-vehicle communication. An artificial neural network was trained offline to predict the lead vehicle’s speed based on road topography and synthetic data. The framework also integrated gear shifting and fuel consumption models. Simulation results demonstrated improved fuel efficiency compared to conventional control strategies. Building on previous efforts that leverage machine learning for fuel estimation, the authors of~\cite{schoen2019machine} introduced a distance-based data summarization method to develop individualized neural network models for heavy-duty vehicles. Unlike traditional time-based approaches, this method aggregated predictors—derived from vehicle speed and road grade—over fixed travel distances. Their results showed that using a 1~km window achieved high prediction accuracy ($R^2 = 0.91$) and less than 4\% mean absolute peak-to-peak percent error across mixed driving conditions. This personalized modeling approach offers a scalable solution for optimizing fuel efficiency across entire fleets.

Although classical models and deep neural networks have proven effective for estimating fuel consumption in individual vehicles, managing and optimizing fuel use across vehicle platoons presents additional complexity. In multi-agent settings, reinforcement learning (RL)~\cite{sutton1999reinforcement} provides a suitable framework for modeling dynamic interactions among vehicles and optimizing long-term fuel efficiency. RL~\cite{ding2020introduction} is a sequential decision-making paradigm in which an agent learns a policy $\pi(a|s)$ that maps states $s \in \mathcal{S}$ to actions $a \in \mathcal{A}$, aiming to maximize the expected cumulative reward over time. Through repeated interaction with the environment, the agent observes transitions $(s_t, a_t, r_t, s_{t+1})$, where $r_t$ is the reward received after taking action $a_t$ in state $s_t$. The learning objective is typically to find an optimal policy $\pi^*$ that maximizes the expected return, defined as $G_t = \mathbb{E} \left[ \sum_{k=0}^\infty \gamma^k r_{t+k} \right]$, where $\gamma \in [0,1]$ is a discount factor~\cite{ding2020introduction}. In multi-agent RL (MARL)~\cite{zhang2021multi}, each vehicle acts as an autonomous agent, and the learning process must account for the non-stationarity introduced by the evolving policies of other agents, making the environment partially observable and more complex. This makes MARL particularly well-suited to cooperative tasks such as platooning, where coordination and adaptability are critical to achieving fuel-efficient behaviors.

The work in~\cite{cunha2022reducing} applied an RL approach to determine a time-switching policy between two target time gap parameters in an Adaptive Cruise Control (ACC) system. Specifically, the authors employed the Proximal Policy Optimization (PPO) algorithm to learn optimal transition timings that minimize fuel consumption in a platoon operating under stochastic traffic conditions. Numerical simulations demonstrated that the PPO-based controller outperformed both fixed time gap ACC and threshold-based switching strategies in terms of average fuel efficiency. While the PPO approach optimized transition timings to minimize fuel consumption, it may struggle with complex platoon dynamics. Therefore, the work in~\cite{li2021reinforcement}, introduced the communication proximal policy optimization (CommPPO) algorithm to handle dynamic platoon configurations like splitting and merging. By using a parameter-sharing structure, CommPPO adapts to varying agent numbers. Compared to existing multi-agent RL strategies and traditional methods, CommPPO achieved an 11.6\% reduction in fuel consumption. In~\cite{zhang2023integrated}, the authors proposed a more integrated approach by addressing both velocity optimization and energy management for hybrid electric vehicle (HEV) platoons. Their approach consisted of a multi-agent RL-based framework that  incorporates energy control strategies and temporal dynamics using Markov games and long short-term memory networks, resulting in a 19.2\% reduction in fuel consumption compared to rule-based methods.

While traditional RL approaches, like those discussed above, perform well in relatively simple environments, Deep Reinforcement Learning (DRL)~\cite{arulkumaran2017deep} extends their capabilities by using deep neural networks to manage complex, high-dimensional state and action spaces. In the following, we review selected studies that apply DRL techniques in this context. For instance, in~\cite{yen2022deep}, the authors developed a DRL framework to manage platoon maneuvers and learn a policy to minimize both fuel consumption and traffic delay by adjusting platoon speed and formation size in response to dynamic traffic conditions. Experimental evaluation demonstrates that the trained DRL agent consistently decreases fuel usage and delay while achieving higher cumulative rewards. In a similar setup, the authors of~\cite{gonccalves2023fuel} proposes a DRL-based switching strategy to minimize fuel consumption in platooning by dynamically selecting between ACC and Cooperative Adaptive Cruise Control (CACC) modes. Simulation results demonstrate that the proposed method surpasses static ACC/CACC and threshold-based controllers in both fuel efficiency and robustness to disturbances. Extending the scope beyond vehicle-level control, the study in~\cite{gao2024drl} integrated traffic signal synchronization and temporal information into the DRL framework to further enhance traffic flow and fuel efficiency. This paper proposed a DRL framework for platooning control that leverages arrival timing vectors (ATVs) and traffic signal synchronization. Experimental results from both corridor and network scenarios demonstrated improvements in average delay and fuel use.

While ML models are inherently “black-box” in nature, they approximate a mapping from inputs such as speed, acceleration, and road gradient to fuel consumption, and are mathematically definable and optimizable during training. However, these models often lack interpretability, and their ability to generalize across different traffic conditions, vehicle types, or dynamic environments is limited. They may struggle to account for the complexities of platooning, such as vehicle interactions, disturbances, or sudden changes in traffic patterns~\cite{matsuo2022deep}. In contrast, RL and, particularly, DRL offer a more flexible and adaptive framework for fuel consumption management in such dynamic environments. By learning control policies through interaction with the environment, these methods allow vehicles to adapt and make decisions based on real-time data, optimizing long-term objectives such as fuel efficiency. In platooning scenarios, RL and DRL techniques can account for varying traffic conditions, vehicle behaviors, and platoon dynamics, continuously improving fuel consumption strategies. Previous studies have highlighted the effectiveness of DRL in learning optimal time-switching policies for ACC and CACC. 

In Table~\ref{tab2}, we provide an overview of the application of machine learning techniques in the literature aimed at reducing fuel consumption. The table highlights various methods, models, and use cases where machine learning has been employed to optimize fuel efficiency in vehicular platoons. These methods have shown significant improvements in fuel efficiency, demonstrating their ability to outperform static or rule-based strategies and adapt to complex, multi-agent environments.

\begin{table*}[htb!]
\caption{Application of Machine Learning Techniques in the Literature for Fuel Consumption Reduction}
\label{tab2}
\begin{tabular}{|p{2cm}|p{4cm}|p{5.3cm}|p{5cm}|}
\hline
\textbf{Category} & \textbf{References} & \textbf{Algorithms} & \textbf{Description} \\ \hline

\textbf{Machine Learning} & 
\begin{tabular}[t]{@{}l@{}}
\makecell{\cite{ekstrom2018estimating} \\ \cite{perrotta2017application}} \\
\makecell{\cite{ping2019impact} \\ \cite{shateri2024utilizing}} \\
\makecell{\cite{xie2023fuel} \\ \cite{hamed2021fuel}} \\
\makecell{\cite{jaffar2020prediction} \\ \cite{moradi2020vehicular}} \\
\makecell{\cite{yang2022predicting}}
\end{tabular} &
\begin{tabular}[t]{@{}l@{}}
• Support Vector Machine (SVM) \\
• Random Forest (RF) \\
• Artificial Neural Network (ANN) \\
• Long Short-Term Memory (LSTM) \\
• Gaussian Process Regression (GPR)
\end{tabular} &
These approaches use data-driven modeling to learn complex relationships between driving conditions and fuel consumption. By training on historical or simulated data, they enable accurate prediction of fuel usage and support optimization strategies for reducing consumption. \\ \hline

\textbf{Reinforcement Learning} & 
\begin{tabular}[t]{@{}l@{}}
\makecell{\cite{qu2020jointly}, \\\cite{CUNHA202299}} \\
\makecell{\cite{berbar2022reinforcement}\\ \cite{chen2023research}} \\
\makecell{\cite{gu2023reinforcement}}
\end{tabular} &
\begin{tabular}[t]{@{}l@{}}
• Deep Deterministic Policy Gradient \\
• Proximal Policy Optimization (PPO) \\
• Q-Learning \\
• Multi-Agent RL (MARL) \\
• Policy Gradient Algorithm
\end{tabular} &
These approaches use RL to develop data-driven control strategies for fuel-efficient vehicle coordination and decision-making. The learned policies are designed to minimize fuel consumption while handling dynamic traffic scenarios. \\ \hline

\textbf{Deep Reinforcement Learning} & 
\begin{tabular}[t]{@{}l@{}}
\makecell{\cite{prathiba2021hybrid}\\ \cite{chen2020intelligent}} \\
\makecell{\cite{lian2023predictive}\\ \cite{kolat2023multi}} \\
\makecell{\cite{irshayyid2023comparative}\\ \cite{liu2020enhancing}} \\
\makecell{\cite{borneo2023platooning}\\ \cite{jiang2022reinforcement}} \\
\makecell{\cite{lu2022sharing}}
\end{tabular} &
\begin{tabular}[t]{@{}l@{}}
• Genetic Algorithm with DRL \\
• Greedy Algorithm + Q-learning \\
• Multi-Agent Soft-Actor-Critic \\
• MARL-PPO \\
• Deep Q-Network (DQN) \\
• Deep Deterministic Policy Gradient
\end{tabular} &
These approaches integrate DRL with heuristic and multi-agent strategies to optimize fuel consumption and coordination in platooning. The use of deep neural networks enables complex policy and value approximations, supporting real-time decision-making in dynamic environments for improved fuel efficiency and cooperative control. \\ \hline
\end{tabular}
\end{table*}

\section{Impact of Platoon Instability on Fuel Consumption}
\label{sec4}
The benefits of platooning are highly dependent on its stability. A platoon is considered stable if disturbances are not amplified as they propagate downstream through the sequence of vehicles~\cite{seiler2004disturbance}. Platoon instability where the disturbance grows exponentially through the platoon is characterized by frequent acceleration and braking among vehicles~\cite{sun2020relationship} can lead to higher fuel consumption due to increased energy expenditure. Therefore, investigating the impact of platoon instability on fuel consumption is a critical task. In the previous section we comprehensively investigated different approaches for fuel consumption estimation in platoons. In the following, we review selected studies from the literature that investigate fuel consumption models, with a focus on their application in evaluating the effects of platoon behavior and instability.

The studies presented in~\cite{knoop2019platoon,ciuffo2021requiem,apostolakis2023energy,he2020energy,wu2011fuel,zhu2019automated,wang2023fuel,shladover2009effects} involve real-world experiments conducted under actual driving conditions, where vehicles with different levels of automation formed platoons on public roads. These experiments primarily aimed to evaluate the influence of ACC systems on traffic dynamics, with a particular emphasis on platoon stability. A key aspect of the research was investigating how instabilities within the platoon—such as fluctuations in speed and spacing—could affect overall fuel consumption. By analyzing vehicle behavior in realistic platooning scenarios, these studies provided valuable insights into the impact of platoon instability on fuel consumption.  

In the same context of platoon instability, other works such as \cite{graffione2020model,sun2019behaviorally,li2016stabilizing,besselink2017string,zheng2022development} focused on developing methodologies to maintain a constant inter-vehicle distance within the platoon, using position and speed reference values as control targets. By reducing the magnitude and frequency of acceleration and braking events, these strategies help prevent string instability and, consequently, reduce fuel consumption. On the other hand, numerous studies have investigated platoon instability from a control-theoretic perspective. For instance, works such as \cite{naus2010string,rajamani2000demonstration,sau2014root,qin2018stability,feng2023controller,li2014stop} focused on analyzing instability of vehicle platoons and designing appropriate control architectures and feedback mechanisms. To this end, control strategies such as CACC, linear feedback controllers, and model predictive control (MPC) have been proposed. A key insight from these studies is that by appropriately tuning the feedback coefficients—particularly those governing position, velocity, and acceleration errors—a balance can be achieved between responsiveness and damping, resulting in improved stability margins leading to less fuel consumption. In a study focused on assistance driving systems, Zhang \textit{et al}.~\cite{zhang2018impact} found that while a simple one-regime Safety Assistance Driving System (SADS) can reduce oscillation magnitude, i.e., eliminating instability, in a car platoon, it may unintentionally increase fuel consumption and emissions.

\section{Open Research Questions}
\label{sec5}
While the reviewed literature provide valuable insights into the mechanisms and approaches for reducing fuel consumption in platooning, several critical limitations and open challenges remain. Identifying these gaps is essential for guiding future research toward more robust, scalable, and real-world-applicable solutions. The following summarizes the key areas where current studies fall short and highlights opportunities for further investigation. 

\begin{itemize}
    \item Most works overlook the role of hybrid electric vehicles (HEVs) within platoons, despite their increasing presence and distinct energy management requirements.
    \item Instability within platoons—such as stop-and-go waves or amplification of small speed changes—also remains a significant challenge, with limited research on its long-term fuel implications.
    \item Although control strategies have been extensively studied, the integration of energy management systems tailored for platooning scenarios is rarely explored. 
    \item A lack of standardized evaluation frameworks hinders direct comparisons across studies.
    \item Most findings rely on simulation-based analyses, with limited large-scale real-world validations under diverse traffic, weather, and road conditions.
    \item Data-driven methods, such as reinforcement learning, are still underdeveloped in this context, particularly in balancing scalability, interpretability, and energy efficiency.
    \item Trade-offs between fuel consumption and other objectives, such as safety, comfort, and communication load, are less investigated.
    \item External factors like road gradient, wind resistance, and vehicle heterogeneity are often oversimplified, leaving room for more realistic modeling efforts.
\end{itemize}
These limitations point to several open directions for future work in order to bridge the gap between simulation and real-world deployment, with the goal to achieve more consistent and optimized fuel consumption in platooning.

\section{Conclusion}
\label{sec6}
This literature review has provided a comprehensive overview of the key factors influencing fuel consumption in vehicle platooning, drawing from mathematical models, experimental findings, and data-driven approaches. Central contributors to fuel efficiency, such as aerodynamic drag reduction, inter-vehicle spacing, and information flow topologies, have been explored in depth. While the literature demonstrates the potential for significant fuel savings, various limitations persist, including a lack of attention to hybrid vehicles, platoon instability, real-time energy management, and validation in real-world, mixed-traffic conditions. These gaps underscore the need for multidisciplinary research and system-level integration to fully realize the benefits of platooning in diverse operational environments. Bridging these challenges will be essential for translating simulation-based gains into sustainable and scalable fuel-efficient transportation solutions.

\section*{Acknowledgment}
This work was conducted while the first author was on internship at the Canadian National Research Council (CNRC). This work was co-funded by Transport Canada’s ecoTECHNOLOGY for Vehicles program and the National Research Council Canada’s Clean and Energy Efficient Transportation program. The views and opinions of the authors expressed herein do not necessarily state or reflect those of Transport Canada.

\bibliographystyle{unsrtnat}




%


\end{document}